\documentclass[twocolumn,aps,nature,preprintnumbers,amsmath,amssymb]{revtex4}
\usepackage{graphicx}  
\usepackage{dcolumn}   
\usepackage{bm}        
\usepackage{amssymb}   
\usepackage[dvipsnames,usenames]{color}
\usepackage{color}
\usepackage{comment}
\usepackage{gensymb}
\usepackage{physics}

\begin{document}
\newcommand{\etal}{{\sl et al.}}
\newcommand{\ie}{{\sl i.e.}}
\newcommand{\sto}{SrTiO$_3$}
\newcommand{\lao}{LaAlO$_3$}
\newcommand{\lno}{LaNiO$_3$}
\newcommand{\nith}{Ni$^{3+}$}
\newcommand{\otw}{O$^{2-}$}
\newcommand{\alo}{Al$_2$O$_3$}
\newcommand{\aalo}{$\alpha$-Al$_2$O$_3$}
\newcommand{\xto}{$X_2$O$_3$}
\newcommand{\eg}{$e_{g}$}
\newcommand{\tg}{$t_{2g}$}
\newcommand{\dzt}{$d_{z^2}$}
\newcommand{\dxtyt}{$d_{x^2-y^2}$}
\newcommand{\dxy}{$d_{xy}$}
\newcommand{\dxz}{$d_{xz}$}
\newcommand{\dyz}{$d_{yz}$}
\newcommand{\egp}{$e_{g}'$}
\newcommand{\ag}{$a_{1g}$}
\newcommand{\mub}{$\mu_{\rm B}$}
\newcommand{\ef}{$E_{\rm F}$}
\newcommand{\alalo}{$a_{\rm Al_2O_3}$}
\newcommand{\asto}{$a_{\rm STO}$}
\newcommand{\nst}{$N_{\rm STO}$}
\newcommand{\lnnlam}{(LNO)$_N$/(LAO)$_M$}
\newcommand{\lxolao}{(La$X$O$_3$)$_2$/(LaAlO$_3$)$_4$}
\newcommand{\lxolno}{(La$X$O$_3$)$_2$/(LaNiO$_3$)$_4$}
\newcommand{\xoalo}{($X_2$O$_3$)$_1$/(Al$_2$O$_3$)$_5$}

\title{Chern and $Z_{2}$ topological insulating phases in perovskite-derived $4d$ and $5d$ oxide buckled honeycomb lattices}

\author{Okan K\"oksal}
\affiliation{Department of Physics and Center for Nanointegration Duisburg-Essen (CENIDE), University of Duisburg-Essen, Lotharstr. 1, 47057 Duisburg, Germany}
\author{Rossitza Pentcheva}
\email{Rossitza.Pentcheva@uni-due.de}
\affiliation{Department of Physics and Center for Nanointegration Duisburg-Essen (CENIDE), University of Duisburg-Essen, Lotharstr. 1, 47057 Duisburg, Germany}
\date{\today}

\begin{abstract}

Based on density functional theory calculations including a Coulomb repulsion parameter $U$, we explore the topological properties of \lxolao(111)
with $X=$ $4d$ and $5d$ cations. The metastable ferromagnetic phases of LaTcO$_3$ and LaPtO$_3$ preserve  P321 symmetry and emerge as Chern insulators (CI) with $C$\,=\,2 and 1 and band gaps of 41 and 38 meV at the lateral lattice constant of LaAlO$_3$, respectively. Berry curvatures, spin textures as well as edge states provide additional insight into the nature of the CI states. While for $X$\,=\,Tc the CI phase is further stabilized  under tensile strain, for $X$\,=\,Pd and Pt a site disproportionation takes place when increasing the lateral lattice constant from $a_{\rm LAO}$ to $a_{\rm LNO}$. The CI phase of  $X$\,=\,Pt shows a strong dependence on the Hubbard $U$ parameter with sign reversal for higher values associated with the change of band gap opening mechanism. Parallels to the previously studied \xoalo\,(0001) honeycomb corundum layers are discussed. Additionally, non-magnetic systems with $X$\,=\,Mo and W are identified as potential candidates for $Z_2$ topological insulators at $a_{\rm LAO}$ with band gaps of 26 and 60 meV, respectively. The computed edge states and $Z_{2}$ invariants underpin the non-trivial topological properties.

\end{abstract}

\maketitle

\section{Introduction}
\label{S:1}

Chern insulators and $Z_{2}$ invariant topological insulators belong to subgroups of topological insulators (TIs) with and without
broken time-reversal symmetry (TRS), respectively.  A Chern insulator, also known as a quantum anomalous Hall insulator (QAHI) exhibits a quantized Hall conductivity without an external magnetic field \cite{Weng2015,Ren2016}. In this context, CI are promising as potential candidates for the realization of Majorana fermions and the application in low-power electronics. The Chern insulator posseses chiral edge states with electrons traversing only in one direction, where the number of conducting edge states is determined by  the Chern number \cite{Thouless}. A $Z_{2}$ invariant TI supports the quantum spin Hall effect (QSHE) which can be regarded as two copies of an IQH (integer quantum Hall) system with electrons forming  Kramers pairs and counter-propogating helical edge states. The properties of QSHE have been addressed explicitly in conjunction with the graphene lattice \cite{Kane2005} and HgTe quantum well structures \cite{Bernevig2006,Zhang2006,Konig2007}. 
Lattices hosting a honeycomb pattern are of particular interest for topologically notrivial states, as initially proposed by Haldane \cite{Haldane}. QAHI have been demonstrated in TIs doped with magnetic impurities such as Mn-doped HgTe or Cr-, Fe-doped Bi$_{2}$Te$_{3}$, Bi$_{2}$Se$_{3}$, Sb$_{2}$Te$_{3}$ \cite{Liu_Zhang,Yu_Zhang,Fang_Bernevig} compounds from the V-VI group. Another possibility to break TRS is by placing $5d$ transition metals on graphene \cite{Zhang2012,Zhou_Liu} or OsCl$_3$ \cite{Sheng2017}. 
Recently, transition metal oxides (TMO) have attracted interest due to their interplay of spin, orbital and lattice degrees of freedom. In contrast to conventional TIs whose bands near the Fermi energy are derived from $s$ and $p$-type orbitals, the narrower $d$-bands lead to larger band gaps and a tendency towards TRS breaking. QAH phases have been predicted both for rocksalt- (EuO/CdO \cite{Zhang2014} and EuO/GdO \cite{Garrity2014}) and  rutile-derived heterostructures \cite{Huang_Vanderbilt,Cai_Gong,Lado2016} and pyrochlore oxides \cite{Fiete2015}. As noticed by Xiao et al. \cite{Xiao2011},   a buckled honeycomb lattice can be formed from two triangular $X$O$_{6}$-layers in the $AX$O$_3$ perovskite structure grown along the [111]-direction. Perovskite-derived bilayers of SrIrO$_3$ and LaAuO$_3$ were proposed as candidates for TIs, however interaction effects in the SrIrO$_3$ bilayer lead to an AFM ground state \cite{Lado2013,Okamoto2014}. $3d$ TM ions tend to host stronger electronic correlations and weaker spin-orbit coupling (SOC). Nevertheless, recently a strong  SOC effect was encountered in (LaMnO$_3$)$_2$/(LaAlO$_3$)$_4$(111)\cite{Doennig2016} which emerges as a Chern insulator with a band gap of ~150 meV when the  symmetry of the two sublattices is constrained. Since the ground state is a Jahn-Teller distorted trivial Mott insulator,  selective excitation of phonons, as recently shown to induce an insulator-to-metal transition in NdNiO$_3$/LaAlO$_3$(001) SLs \cite{caviglia12}, may present a pathway to  suppress the symmetry breaking and access the CI state. 
Alternatively, $4d$ and $5d$ systems turn out to be less sensitive to symmetry breaking and the interplay of weaker correlation and stronger SOC makes them interesting candidates. This design principle served to identify LaRuO$_3$ and LaOsO$_3$ honeycomb bilayers sandwiched in LaAlO$_3$(111) as Chern insulators \cite{HongliNQM}. Both Ru$^{3+}$ ($4d^5$) and Os$^{3+}$ ($5d^5$) are in the low-spin state with a single hole in the \tg\ manifold whereas the homologous Fe$^{3+}$  ($3d^5$)  in LaFeO$_3$ is found to be in a high-spin state with an AFM ground state.

\begin{figure}[h!]
\centering
\includegraphics[width=9.2cm,keepaspectratio]{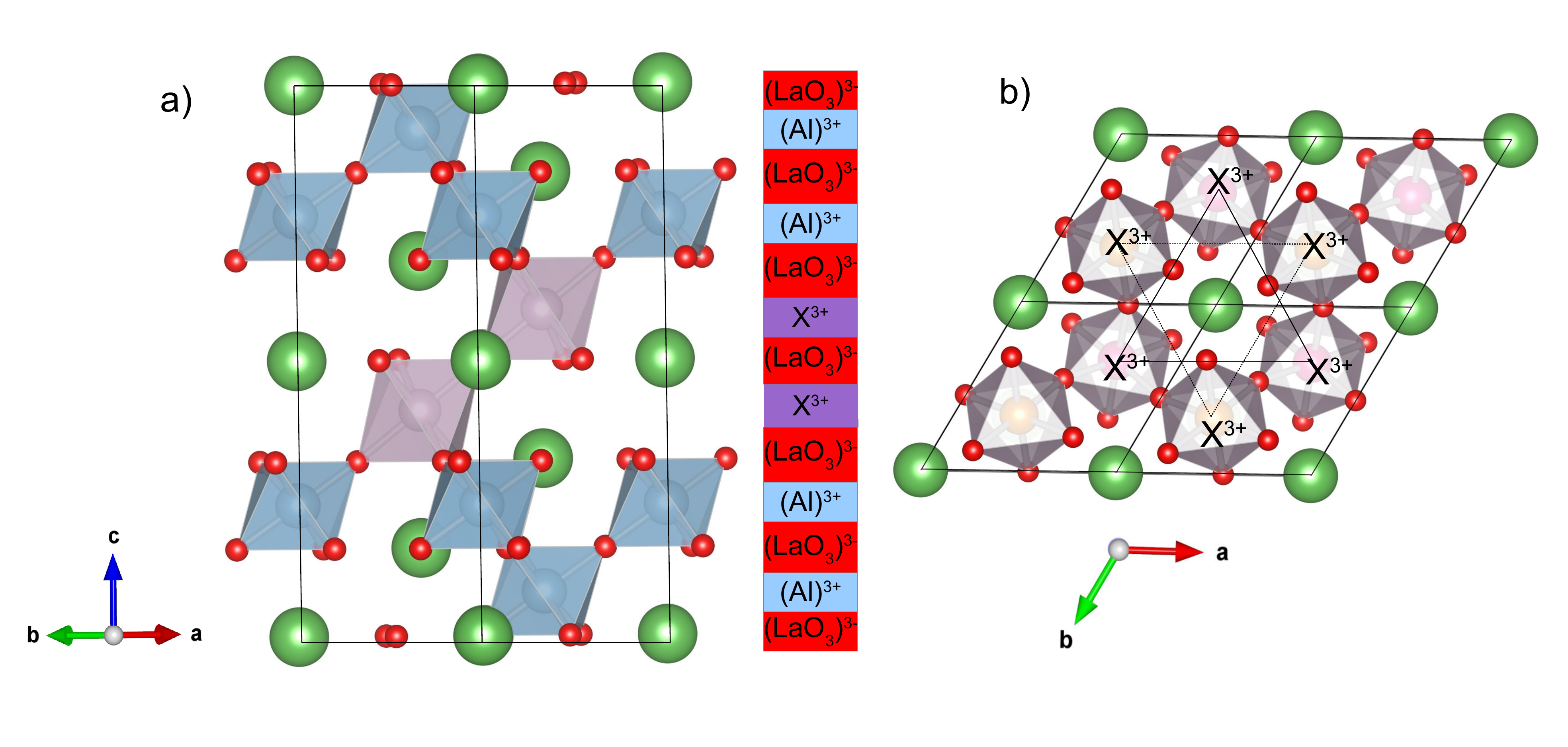}
\caption{a) Side view of the \lxolao(111) superlattice where $X$ represents a $4d$, $5d$ cation. b) Top view of the buckled honeycomb lattice in the $a$-$b$ plane illustrating the corner-sharing octahedra in the perovskite whereas solid and dashed lines connect the next nearest TM-ion neighbors residing on the same sublattices.} 
\label{fig:schematic_view}
\end{figure}

A honeycomb pattern arises also in the corundum structure, albeit with smaller buckling and different connectivity. While in the perovskite structure the octahedra are corner sharing (cf. Fig. \ref{fig:schematic_view}), in the corundum-derived SLs the $X$O$_6$ octahedra in the $X_2$O$_3$ layer are edge-sharing as well as alternating corner- and face-sharing to the next layer above and below. The complex electronic behavior of corundum-derived honeycomb layers \xoalo(0001) were recently addressed in a systematic study of the $3d$ series \cite{OKRP2018}. Moreover, in the $4d$ and $5d$ systems the ferromagnetic cases of $X$\,=\,Tc,\,Pt were identified as Chern insulators with $C$\,=\,-2 and -1 and band gaps  of 54 and 59 meV, respectively\cite{JPCS2018}. 
This motivated us to explore here the perovskite analogues $X$\,=\,Tc, Pd, and Pt in (111)-oriented \lxolao. Although the ground state is AFM, we find that a CI phase emerges  for the metastable FM cases with $C$\,=\,2 and 1. Furthermore, we explore the effect of strain on the stability of the CI state, as well as the dependence on the Coulomb repulsion parameter $U$ 
and compare to the corundum-type systems. 
Last but not least, we concentrate on TI cases where time reversal and inversion symmetry are preserved and identify the non-magnetic phases of $X$\,=\,Mo,\,W in \lxolao(111) superlattices as potential candidates for $Z_2$ TIs. 

\section{Theoretical methods}

Density functional theory calculations were performed for \lxolao(111) SLs employing the projector augmented wave (PAW) method \cite{PAW} as implemented in the VASP \cite{VASP} code. The plane-wave cutoff energy is fixed to 600 eV. The generalized gradient approximation (GGA) in the parametrization of Perdew-Burke-Enzerhof \cite{GGA_PBE} was used for the exchange-correlation functional.
The static local correlation effects were accounted for in the GGA\,+$U$ approach, using  $U_{\rm eff}=U-J$ of Dudarev et al.\cite{Dudarev}. Hubbard $U$ values for the $4d$ and $5d$ ions are typically lower than for the $3d$ cations \cite{HongliNQM,JPCS2018}. We used an $U$\,=\,3 eV for $X$\,=\,Tc,\,Pd,\,Mo and 1-2 eV for $X$\,=\,Pt,\,W with a Hund's exchange parameter of $J$\,=\,0.5 eV on all cations. Additionally, $U=8$ eV is used for the empty  $4f$ orbitals of La. 
The calculations were performed using a $\Gamma$ centered $k$-point grid of 12$\times$12$\times$2. The lattice parameter $c$ and the internal coordinates of the superlattice structure were optimized until the Hellman-Feynman forces were less than 1 meV/\AA. SOC was considered in the second-variational method with magnetization along the (001) quantization axis. For potentially interesting cases maximally localized Wannier functions (MLWFs) were constructed in order to calculate the Berry curvatures and the anomalous Hall conductivity (AHC) on a dense $k$-point mesh of 144$\times$144$\times$12 using the wannier90 code \cite{wannier90}.

\section{Results and discussion}
\label{sec:results_discussion}

Our previous studies \cite{Doennig2016,HongliNQM,OKRP2018,JPCS2018} on systems hosting the honeycomb lattice with $X=4d$ and $5d$ showed that the metastable ferromagnetic state of $X$\,=\,Tc,\,Pt host quantum anomalous Hall states. 
Using the insight gained from this investigation, we explore the perovskite-derived SLs with the above-mentioned TM ions. Although the ground states of $X$\,=\,Tc,\,Pd and Pt in \lxolao(111) superlattices are AFM, (cf. Table~\ref{tab:latparam_bilayer_Tc_Pd_Pt}) with symmetry lowering due to a dimerization, manifested in alternating $X-X$ bond lengths (not shown here), we concentrate here  on the metastable ferromagnetic phases and explore their topological properties.  Moreover, we investigate the effect of strain on the Chern insulating phases by considering two in-plane lattice constants of substrates LaAlO$_3$ (3.79~\AA) and LaNiO$_3$ (3.86~\AA) with a relative strain of $\sim1.8$\%.
We futhermore extend our study to non-magnetic solutions leading to $Z_2$ TIs. In particular, $X$\,=\,Mo,\,W turn out to possess $Z_{2}$ topologically invariant phases, albeit their  non-magnetic phases are higher in energy by 2.0 eV and 0.4 eV per u.c., respectively, compared to the antiferromagnetic ground states (cf. Table~\ref{tab:latparam_bilayer_Mo_W}). 

\begin{table}[ht]
\centering
\caption{Structural, electronic and magnetic properties of the FM state in $X$\,=\,Tc,\,Pt,\,Pd SLs at $a_{\rm LAO}$ and $a_{\rm LNO}$, respectively. 
The relative energy difference of ferromagnetic (FM) configuration with respect to the antiferromagnetic (AFM) ground state is given. Spin and orbital moments are in \mub. The corresponding band gaps with and without SOC are given in meV ($"m"$ denotes a metallic state). For each of the two lattice constants, the resulting Chern numbers are also listed.}
\label{tab:latparam_bilayer_Tc_Pd_Pt}
\begin{tabular}{>{\raggedright}p{0.28\linewidth}>{\raggedright}p{0.22\linewidth}>{\raggedright\arraybackslash}p{0.16\linewidth}} \toprule
2LTcO & $a_{\rm LAO}$ & $a_{\rm LNO}$ \\ \hline
$c$[\AA] & 13.9 & 13.8 \\ 
$\Delta E_{\rm FM-AFM}$[eV]  &1.0 &--   \\ 
$M_S$[\mub] &1.96/1.96 &1.96/1.96  \\ 
$M_L$[\mub] &0.06/0.06 &0.06/0.06  \\ 
E$_g$[meV]  &m &m  \\ 
E$_{g(\rm SOC)}$[meV]  &41 &53  \\ 
$C$ &2 &2  \\ \hline 

2LPdO & $a_{\rm LAO}$ & $a_{\rm LNO}$ \\ \hline
$c$[\AA] & 14.0 & 13.7 \\ 
$\Delta E_{\rm FM-AFM}$ [eV] &0.8 &--  \\ 
$M_S$[\mub] &0.71/0.72 &0.60/0.85  \\ 
$M_L$[\mub] &0.05/0.05 &0.05/0.07  \\ 
E$_g$[meV]  &m &60  \\ 
E$_{g(\rm SOC)}$[meV]  &m &70  \\ 
$C$ &0 &0  \\ \hline 

2LPtO & $a_{\rm LAO}$ & $a_{\rm LNO}$ \\ \hline
$c$[\AA] & 14.2 & 14.1 \\ 
$\Delta E_{\rm FM-AFM}$[eV] &1.1 &--   \\ 
$M_S$[\mub] &0.67/0.67 &1.06/0.32  \\ 
$M_L$[\mub] &0.13/0.13 &0.25/0.05  \\ 
E$_g$[meV]  &m &370  \\ 
E$_{g(\rm  SOC)}$[meV]  &38 &402  \\ 
$C$ &1 &0  \\ \hline \hline
\end{tabular}
\end{table}

\subsection{GGA+$U$(+SOC) results: $X$\,=\,${\rm Tc}$}
\label{sec:Tc_electronic_and_topological_properties}

\begin{figure} [htbp!]
\includegraphics[width=9.2cm,keepaspectratio]{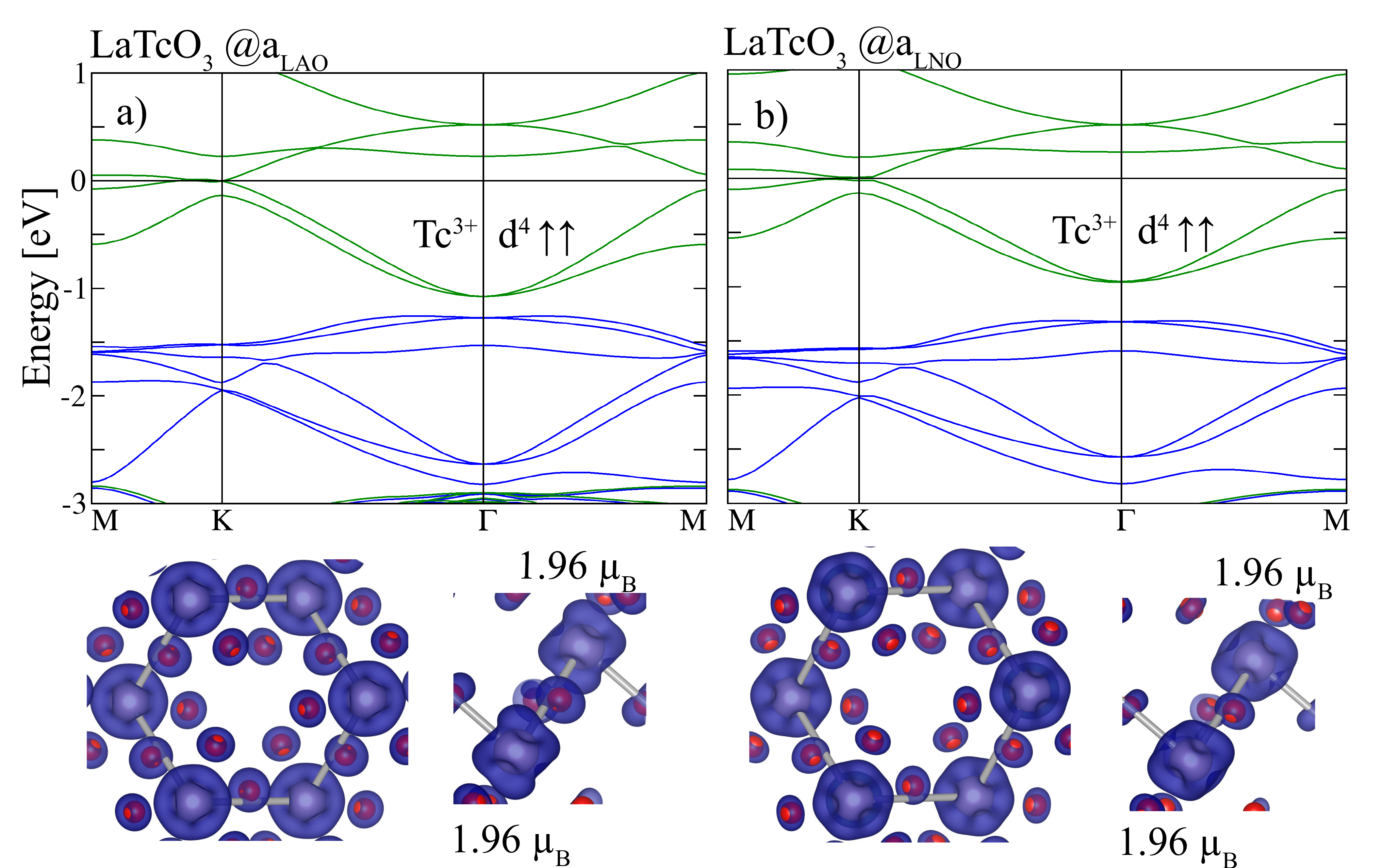}
\caption{Spin-resolved band structure of the buckled bilayers for $X$\,=\,Tc with corresponding lattice constants $a$ at a) $a_{\rm LAO}$ and b) $a_{\rm LNO}$, respectively. In the band structures blue/green denote majority/minority bands and the Fermi level is set to zero. The lower two panels depict the top and side view of isosurfaces of the spin densities in the integration range of --3 eV to \ef\, where  blue (red) show the  majority (minority) contributions.}
\label{fig:Tc_LAO_LNO}
\end{figure}

In the following, we discuss the electronic and topological properties of ferromagnetic $X$\,=\,Tc for in-plane lattice constants $a_{\rm LAO}$ and $a_{\rm LNO}$. Without SOC and for $U_{\rm eff}$\,=\,2.5 eV semi-metallic band structures emerge as depicted in Fig. \ref{fig:Tc_LAO_LNO}a and b. In both cases the band structures around \ef\ are very similar and dominated by minority Tc \tg\ bands  (cf. Fig. \ref{fig:Tc_LAO_LNO}a-b) touching at K, that extend to $\sim-1.4$ eV ($a_{\rm LAO}$) and $\sim-1.0$ eV ($a_{\rm LNO}$) and are completely separated from the lower lying majority bands. This feature is dissimilar to the band structure in the corundum honeycomb layer (Tc$_{2}$O$_{3}$)$_{1}$/(Al$_{2}$O$_{3}$)$_{5}$(0001) \cite{JPCS2018} where the majority and minority bands are entangled around \ef\,. 

The  spin densities are shown in \ref{fig:Tc_LAO_LNO}a and b. In the  Tc$^{3+}$ $4d^4$ configuration   all electrons are in the \tg\ subset with two unpaired electrons, reflected in a magnetic moment of 1.96 \mub\ for both Tc sites. This is in contrast to the corundum honeycomb layer with $X$\,=\,Tc \cite{JPCS2018} where a much lower magnetic moment of 0.93~\mub\ is found resulting from a violation of Hund's rule due to a strong hybridization between Tc $4d$ and O $2p$ states. 

\begin{figure} [htbp!]
\centering
\includegraphics[width=9.2cm,keepaspectratio]{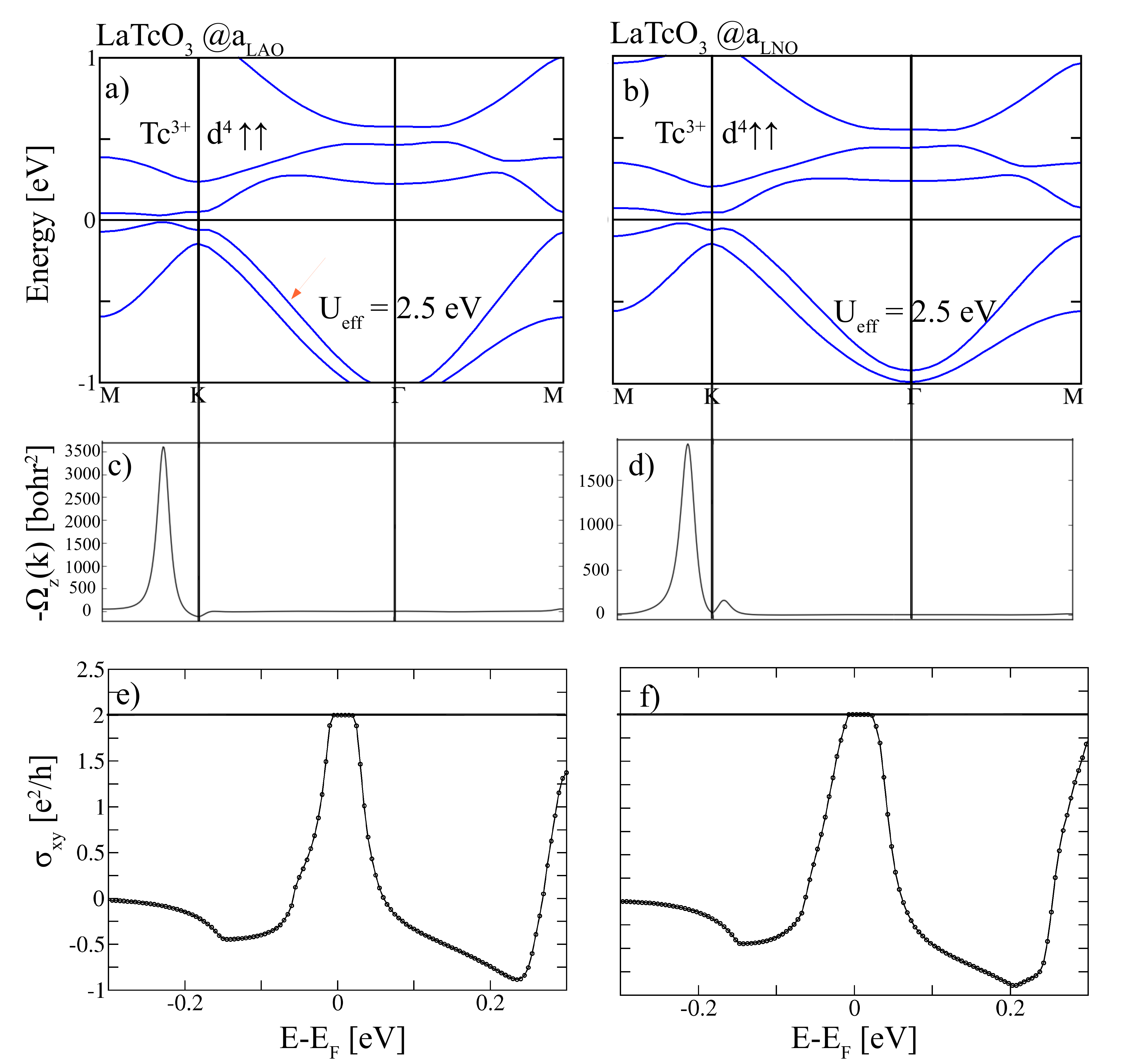}
\caption{GGA\,+\,\textit{U}\,+\,SOC band structures for $X$\,=\,Tc in (111)-oriented perovskite bilayer at a) $a_{\rm LAO}$ and b) $a_{\rm LNO}$. The Berry curvatures (c-d) are plotted along the same $\textit{k}$-path and (e-f) show the corresponding anomalous Hall conductivities $\sigma_{xy}$  vs. the chemical potential in units of $e^{2}/h$.}
\label{fig:Tc_LAO_LNO_SOC}
\end{figure}

We proceed with the effect of SOC on the band structures illustrated in Fig. \ref{fig:Tc_LAO_LNO_SOC}a and b. The strongest influence is observed at the Fermi level around the K point where SOC induces anti-crossings and opens gaps of 41 and 53 meV for $a_{\rm LAO}$ and $a_{\rm LNO}$, respectively. Calculation of the AHC (Fig. \ref{fig:Tc_LAO_LNO_SOC}e-f) shows that the LaTcO$_3$ honeycomb layer emerges as Chern insulator with $C$\,=\,2 for both strain values. The largest contributions to the Berry curvature $\Omega(k)$  (Fig. \ref{fig:Tc_LAO_LNO_SOC}c-d) arises along K-M. The enhanced gap for $a_{\rm LNO}$ leads to a broader Hall plateau at \ef\ in Fig. \ref{fig:Tc_LAO_LNO_SOC}e and f. 
The results demonstrate that with tensile strain the CI phase is further stabilized. A similar effect of strain was observed in (Tc$_{2}$O$_{3}$)$_{1}$/(Al$_{2}$O$_{3}$)$_{5}$(0001) \cite{JPCS2018}. We note that the sign of the Chern number $C$\,=\,2 for $X$\,=\,Tc in the (111)-oriented perovskite bilayer is reversed compared to the corundum-derived SL ($C$\,=\,-2) \cite{JPCS2018}. {\bf The reversal of sign is related to the specific band topology and band gap opening mechanism and the predominance of minority bands, whereas in the  corundum case  majority bands reside around \ef \cite{JPCS2018}}.

\subsection{GGA+$U$ results for isoelectronic $X$\,=\,${\rm Pd}$,\,${\rm Pt}$}
\label{sec:Pd_Pt_electronic_and_topological_properties}

\begin{figure} [htbp!]
\centering
\includegraphics[width=9.2cm,keepaspectratio]{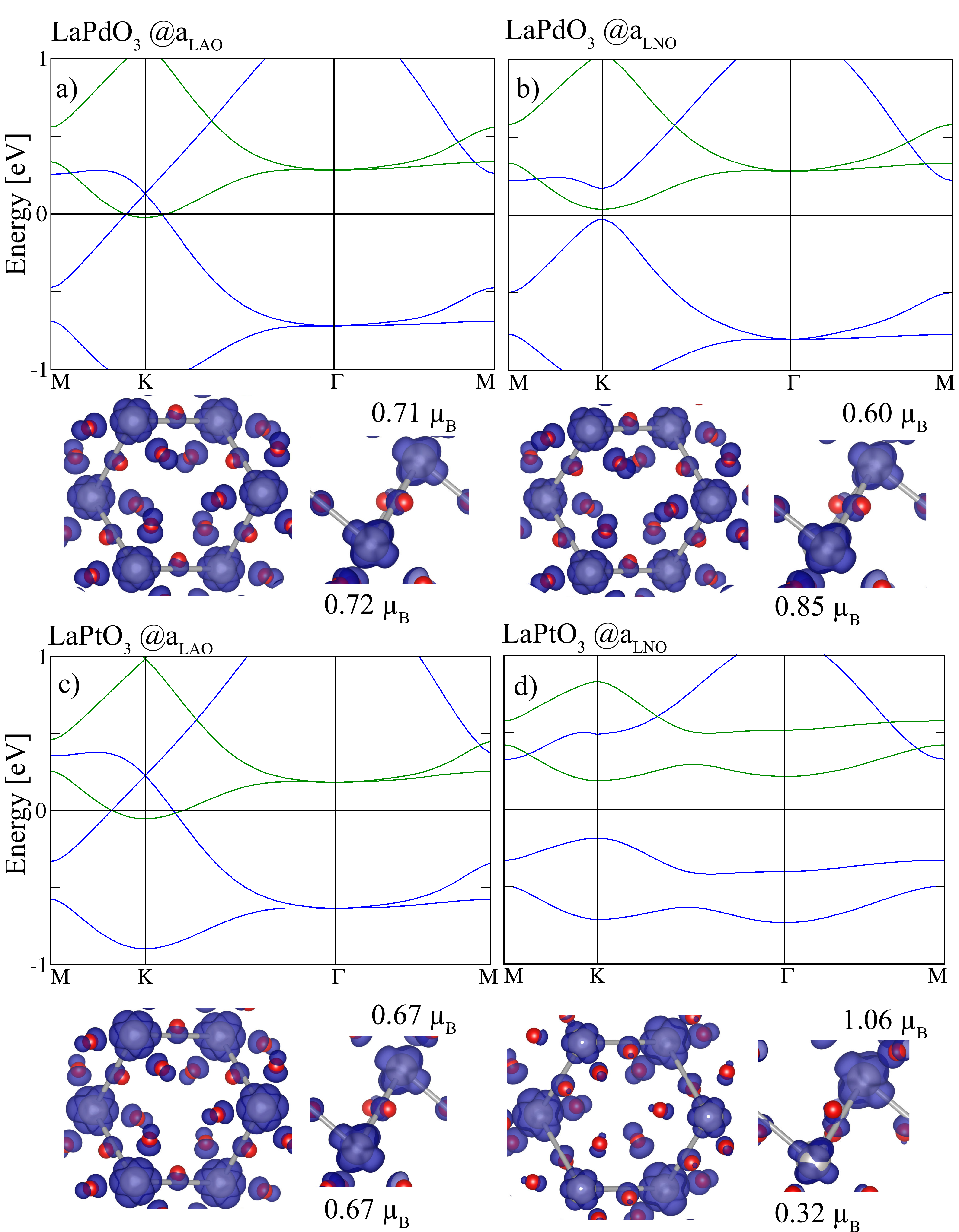}
\caption{Spin-resolved band structure of the buckled bilayers in ((111)-oriented \lxolao\ perovskite bilayer (a-c) at $a_{\rm LAO}$ and (b-d) $a_{\rm LNO}$ for the isovalent and isoelectronic $X$\,=\,Pd and Pt at $U_{\rm eff}$\,=\,2.5 eV and 1.0 eV. The top and side view of isosurfaces of the spin densities are integrated in the energy range between --1 (Pt) and --1.2 eV (Pd) to \ef\ where  blue (red) show the  majority (minority) contributions.}
\label{fig:Pd_Pt_LAO_LNO}
\end{figure}

We next turn to the isoelectronic $X$\,=\,Pd and Pt. Experimental studies suggest paramagnetic metallic behavior for  bulk LaPdO$_3$ \cite{Kim2001,Kim2002}.  For the honeycomb layer of $X$\,=\,Pd and Pt at $a_{\rm LAO}$  $P321$ symmetry is preserved and the band structures  in Fig. \ref{fig:Pd_Pt_LAO_LNO}a and c show two very similar sets of each four majority and minority bands, the former lying about 1 eV lower than the latter. Both exhibit Dirac crossings at K, the one of the majority band being slightly above the Fermi level. Consequently, the dispersive majority and the bottom of the minority bands cross $E_{\rm F}$ and lead to a metallic state.  The situation is thus somewhat different from the isoelectronic LaNiO$_3$ analogon where for $P321$ symmetry the Dirac point is fixed at the Fermi level.\cite{Fiete2011,Yang2011,Doennig2014,Doennig2016}  
A substantial occupation of both \eg\ orbitals and contribution from the O $2p$ states is visible from the spin-densities (see Fig. \ref{fig:Pd_Pt_LAO_LNO}) which indicates a $d^8L$ configuration  instead of the formal $d^7$ occupation and bears analogies to the isoelectronic LaNiO$_3$ \cite{Freeland2011,BlancaRomero2011,Doennig2014}.  In contrast, for tensile strain ($a_{\rm LNO}$) the P321 symmetry is lowered and a gap of $\sim 60$ meV and $\sim 370$ meV (cf. Fig. \ref{fig:Pd_Pt_LAO_LNO}b and d) is opened for LaPdO$_3$ and LaPtO$_3$, respectively. The gap opening arises due to the disproportionation of the two Pd and Pt triangular sublattices expressed in different magnetic moments: In $X$=\,Pd the two sites acquire magnetic moments of 0.85~\mub\ and 0.60~\mub\ (cf. Fig. \ref{fig:Pd_Pt_LAO_LNO}b). For $X$=\,Pt this site-disproportionation is more pronounced with magnetic moments of 1.06~\mub\ and 0.32~\mub\ on the two Pt sites (cf. Fig. \ref{fig:Pd_Pt_LAO_LNO}d) resulting in a larger gap between the occupied majority and unoccupied minority pairs of bands whose dispersion is significantly reduced. The site-disporportionation at tensile strain goes hand in hand with a breathing mode by showing a larger and a smaller PtO$_6$ octahedron with volumes 13.8 and 11.8~\AA$^3$, respectively. As a consequence, the Pt-O bond lengths result in 2.17, 2.18 \AA\ at the first and 2.03, 2.10 \AA\ at the second Pt site. Such a disproportionation is common in bulk rare earth nickelates\cite{millis,sawatzky}, (001) or (111)-oriented LaNiO$_3$/LaAlO$_3$ SLs \cite{Boris2011,Freeland2011,BlancaRomero2011,Geisler2018,Doennig2014,Doennig2016} as well as La$_2$CuO$_4$/LaNiO$_3$(001) SLs\cite{wrobel2018}.

\subsection{$X$\,=\,${\rm Pt}$: Emergence of a CI phase as a function of $U$}
\label{sec:Pt_CI_phases}

\begin{figure*} [htbp!]
\includegraphics[width=18cm,keepaspectratio]{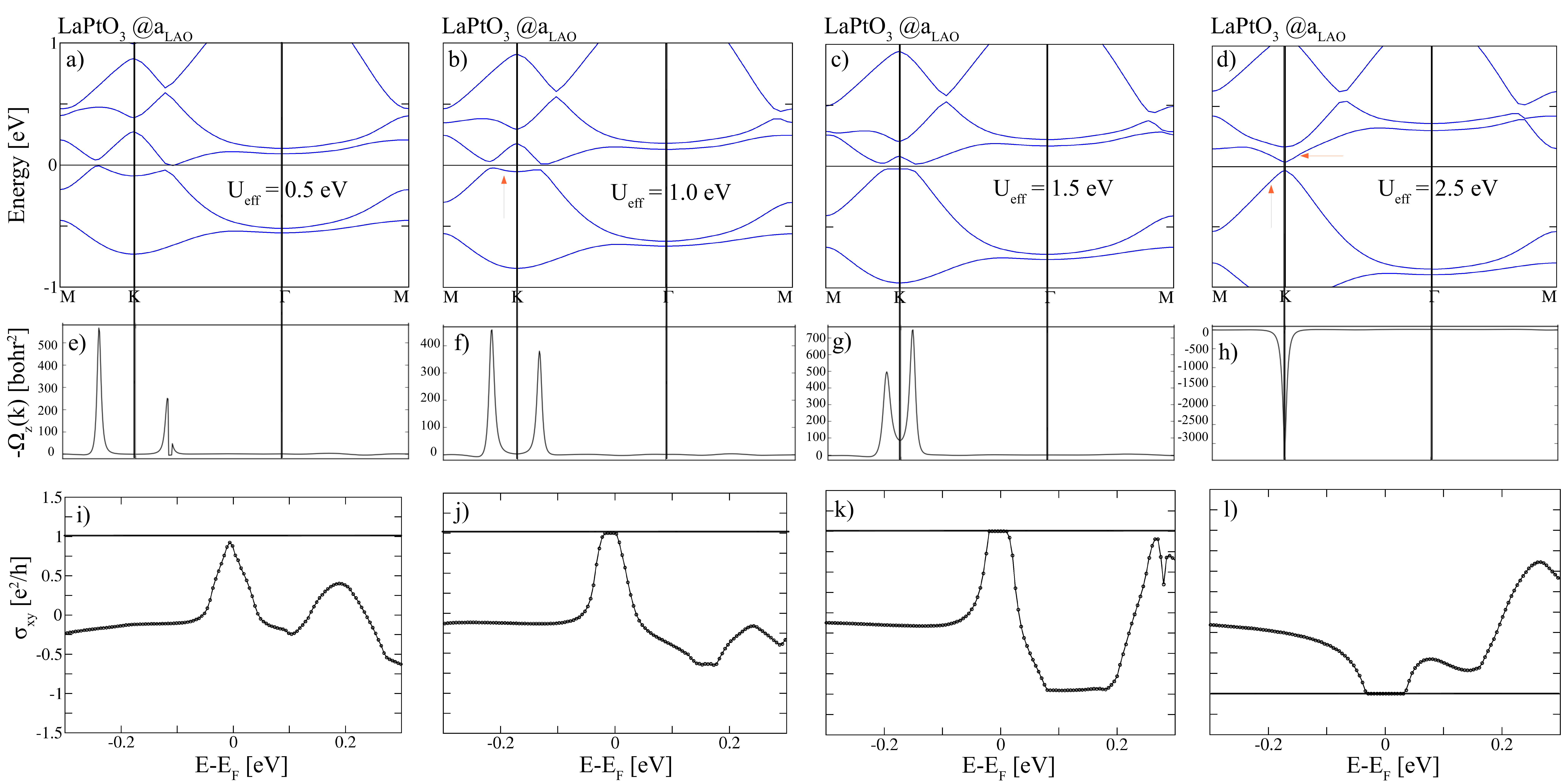}
\caption{GGA\,+\,\textit{U}\,+\,SOC results for $X$\,=\,Pt in (111)-oriented \lxolao\ as a function of the Coulomb repulsion parameter $U$: 
Evolution of the band structure (a-d), Berry curvatures (e-h) plotted along  the same $\textit{k}$-path  and AHC $\sigma_{xy}$ vs. the chemical potential (i-l).}
\label{fig:Pt_BC_AHC}
\end{figure*}

Since for $a_{\rm LNO}$ both LaPdO$_3$ and LaPtO$_3$ result in trivial Mott insulators due to site disproportionation, we explore here the topological properties at $a_{\rm LAO}$, where the P321 symmetry is preserved. Upon including SOC, for $X$\,=\,Pd a CI phase emerges for $U$ values beyond $U_{\rm eff}^{\rm c}$\,=\,3.5 eV (not shown here), which are likely too high for a $4d$ element. We concentrate here on the topological properties of (LaPtO$_3$)$_2$/(LaAlO$_3$)$_4$(111) as a function of Hubbard $U$. Up to $U_{\rm eff}$\,=\,2.0 eV  SOC leads to a band inversion between the majority and minority bands around K. At $U_{\rm eff}$\,=\,0.5 eV the conduction band still overlaps with the Fermi level (cf. Fig. \ref{fig:Pt_BC_AHC}a) and hampers the formation of a quantized Hall plateau (see Fig. \ref{fig:Pt_BC_AHC}i). For 1.0$<$$U_{\rm eff}$$<$2.0 eV (cf. Fig. \ref{fig:Pt_BC_AHC}b and c), the Fermi level lies inside the gap and the system becomes a Chern insulator with $C$\,=\,1.  Increasing the Coulomb repulsion strength from 1.0 to 1.5 eV enhances the band gap (from 31 to 38 meV) and the Hall plateau which stabilizes the Chern insulating phase (see Fig. \ref{fig:Pt_BC_AHC}k). As can be seen from Figures \ref{fig:Pt_BC_AHC}f and g, positive contributions to the Berry curvature arise around K. For higher values $U_{\rm eff}$\,$\geq$\,2.5 eV the effect of SOC changes from band inversion between bands of opposite spin to avoided crossing  between two bands with a larger gap of 66 meV, leading to a sign reversal  of the Chern number  from +1 to --1 (cf. Fig. \ref{fig:Pt_BC_AHC}l). This is consistent with the large negative Berry curvature contribution around K in Fig. \ref{fig:Pt_BC_AHC}h. However, for $5d$ systems $U$ values beyond 2.0 eV appear to be too high to describe correctly the electronic properties.

\subsection{Edge states and spin textures of CI phases}
\label{sec:Spin_textures_CI_phases}

\begin{figure} [htbp!]
\centering
\includegraphics[width=9.2cm,keepaspectratio]{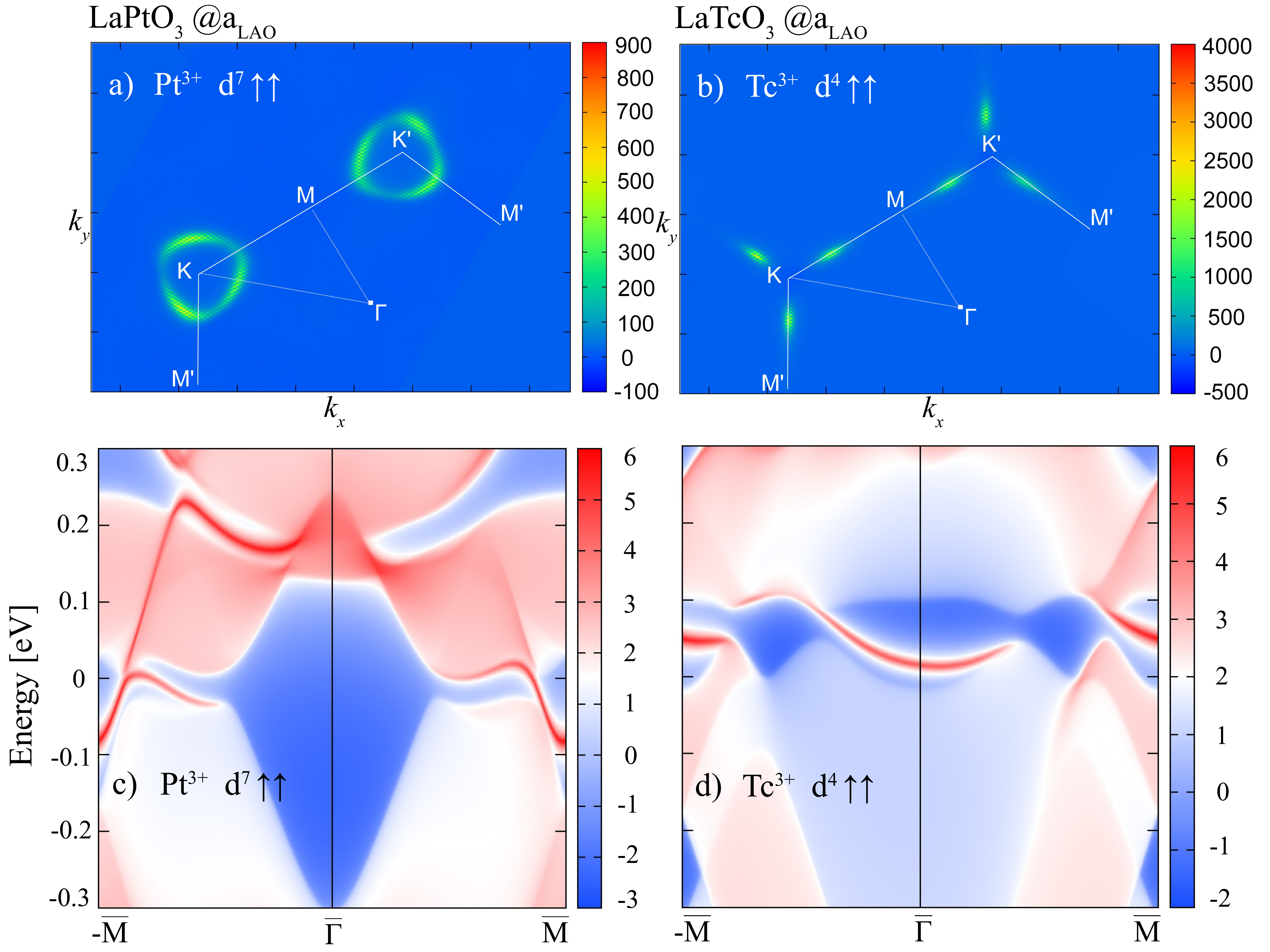}
\caption{Top view of the Berry curvatures $\Omega(k)$ for a) $X$\,=\,Pt b) Tc in (111)-oriented \lxolao\ at $U_{\rm eff}$\,=\,1.0 eV and 2.5 eV in the Chern-insulating phase. The calculated edge states $X$\,=\,Pt,\,Tc superlattices shown in (c-d) for (100) surfaces. Red-white range of colors represent higher local DOS, the solid red lines correspond to the edge states connecting valence and conduction bands. The blue regions denote the bulk energy gap. The Fermi level is set to zero.}
\label{fig:Tc_Pt_BC_edge_states}
\end{figure}

\begin{figure*} [htbp!]
\includegraphics[width=12cm,keepaspectratio]{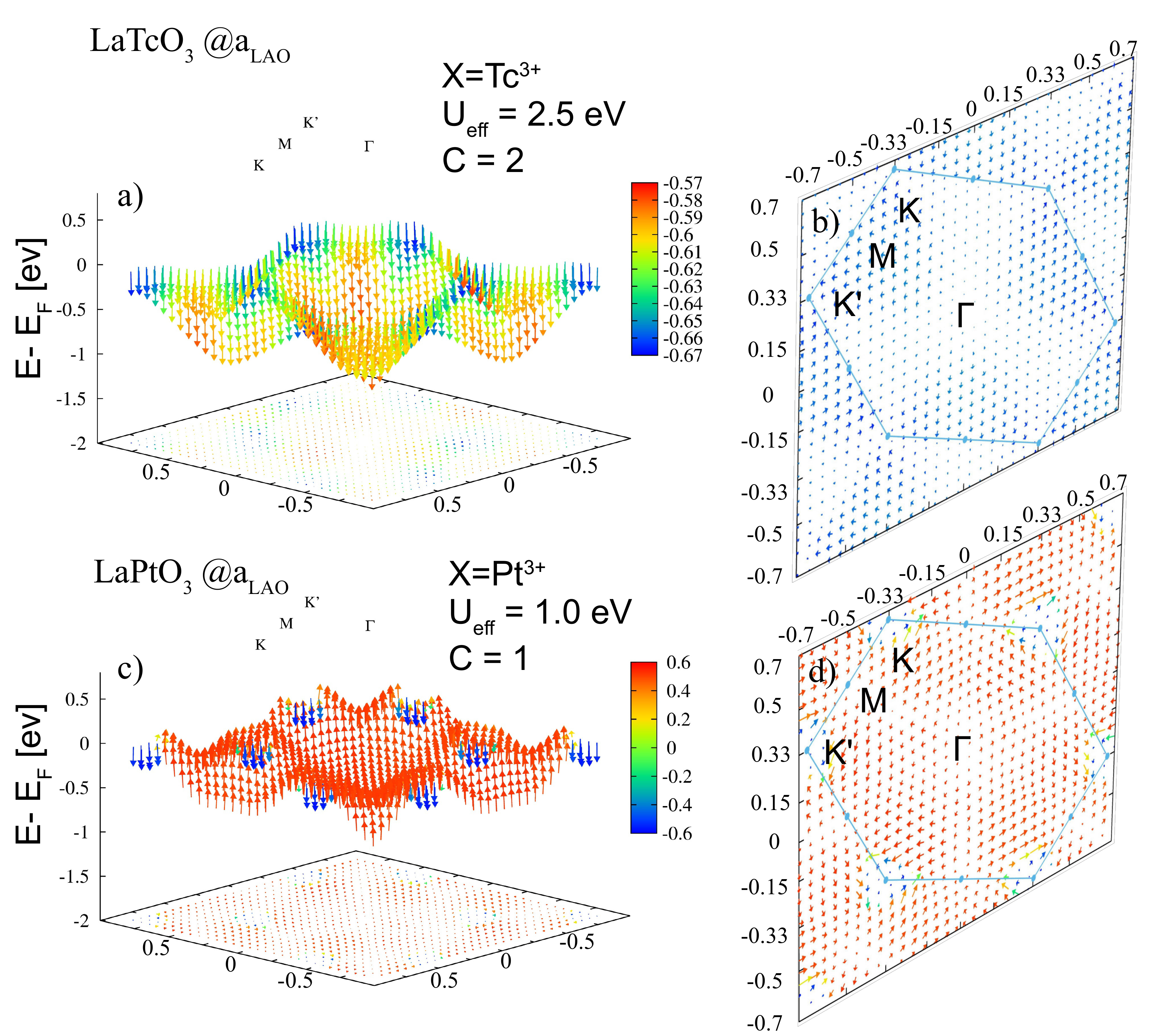}
\caption{Side and top view of the spin textures in $k$-space from GGA\,+\,\textit{U}\,+\,SOC calculations with out-of-plane magnetization of the highest occupied bands for (a-b) $X$\,=\,Tc (cf. Fig. \ref{fig:Tc_LAO_LNO_SOC}a) at $U_{\rm eff}$\,=\,2.5 eV and (c-d) for $X$\,=\,Pt (cf. Fig. \ref{fig:Pt_BC_AHC}b) at $U_{\rm eff}$\,=\,1.0 eV. The color scale provides the projection of the texture field on the $\hat{z}$-axis with red (blue) indicating parallel (antiparallel) orientation for $X$\,=\,Pt. In contrast, for $X$\,=\,Tc only negative values with varying size are observed. The top views display the in-plane variation of the spins.}
\label{fig:Spin_texture_Tc_Pt}
\end{figure*}

Unlike in (Pt$_{2}$O$_{3}$)$_{1}$/(Al$_{2}$O$_{3}$)$_{5}$(0001) \cite{JPCS2018} where contributions to the Berry curvature $\Omega(k)$ arise along M and K, the top view of the Berry curvature for $U_{\rm eff}$\,=\,1.0 eV (see Fig. \ref{fig:Tc_Pt_BC_edge_states}a) reveals that non-vanishing contributions appear solely on a rounded triangular feature around K marking the anticrossing line  of the majority and minority band. The surface state in Fig. \ref{fig:Tc_Pt_BC_edge_states}c calculated  employing the MLWF method \cite{wanniertools} presents a single chiral edge state associated with  $C$\,=\,1. For $X$\,=\,Tc the largest contribution to $\Omega(k)$ emerges along K-M (cf. Fig. \ref{fig:Tc_Pt_BC_edge_states}b and d) resulting in two in-gap chiral states whose features are similar to (Tc$_{2}$O$_{3}$)$_{1}$/(Al$_{2}$O$_{3}$)$_{5}$(0001) \cite{JPCS2018}.

Here we briefly address the spin texture  of the highest occupied  band (marked by arrows in Fig. \ref{fig:Tc_LAO_LNO_SOC}a and Fig. \ref{fig:Pt_BC_AHC}b) in the Chern insulating phase for $X$\,=\,Tc and $X$\,=\,Pt, respectively. For $X$\,=\,Pt (see Fig. \ref{fig:Spin_texture_Tc_Pt}c) the spin texture is dominated by majority (red) components in the larger part of the BZ and exhibits  an orientation reversal of minority (blue) $s_z$ spin components close to K, consistent with the SOC-induced band inversion between the occupied majority and unoccupied minority band around K discussed above. The spin texture of (LaTcO$_3$)$_2$/(LaAlO$_3$)$_4$(111) in Fig. \ref{fig:Spin_texture_Tc_Pt}a is rather collinear compared to the corundum case which shows a vortex  around $\Gamma$. In contrast, the spin texture  for $X$\,=\,Tc in Fig. \ref{fig:Spin_texture_Tc_Pt}a exhibits only negative $s_z$ values throughout the entire BZ. This is in agreement with the fact that only bands of minority character appear around \ef. Overall, even though the number of edge states of the perovskite and corundum case (Tc$_{2}$O$_{3}$)$_{1}$/(Al$_{2}$O$_{3}$)$_{5}$(0001) \cite{JPCS2018} are identical, the differences can be attributed to the distinct electronic structure and the effect of SOC, as discussed in Section \ref{sec:Tc_electronic_and_topological_properties}.

\subsection{$Z_{2}$ topological invariant systems: GGA+$U$(+SOC) results for isoelectronic $X$\,=\,${\rm Mo}$,\,${\rm W}$}
\label{sec:Mo_W_electronic_and_topological_properties}

\begin{table}[ht]
\centering
\caption{Structural, electronic and magnetic properties of $X$\,=\,Mo,\,W SLs in \lxolao(111). The relative energy difference of the  non-magnetic (NM) configuration with respect to the antiferromagnetic (AFM) ground state is displayed. The corresponding band gaps with and without SOC are given in meV, respectively. The $Z_{2}$ indices are also listed.}
\label{tab:latparam_bilayer_Mo_W}
\begin{tabular}{>{\raggedright}p{0.28\linewidth}>{\raggedright}p{0.22\linewidth}>{\raggedright\arraybackslash}p{0.14\linewidth}} \toprule
&2LMoO 		&2LWO \\ \hline   	 
&$a_{\rm LAO}$  &$a_{\rm LAO}$  \\ \hline
$c$[\AA] & 14.1  & 14.2 \\ 
$\Delta E_{\rm NM-AFM}$[eV] &2.0  &0.4 \\ 
E$_g$[meV]  &26  &67 \\ 
E$_{g(\rm SOC)}$[meV]  &26  &60 \\ 
$Z_{2}$ &1 &1  \\ \hline \hline
\end{tabular}
\end{table}

\begin{figure} [htbp!]
\includegraphics[width=9.2cm,keepaspectratio]{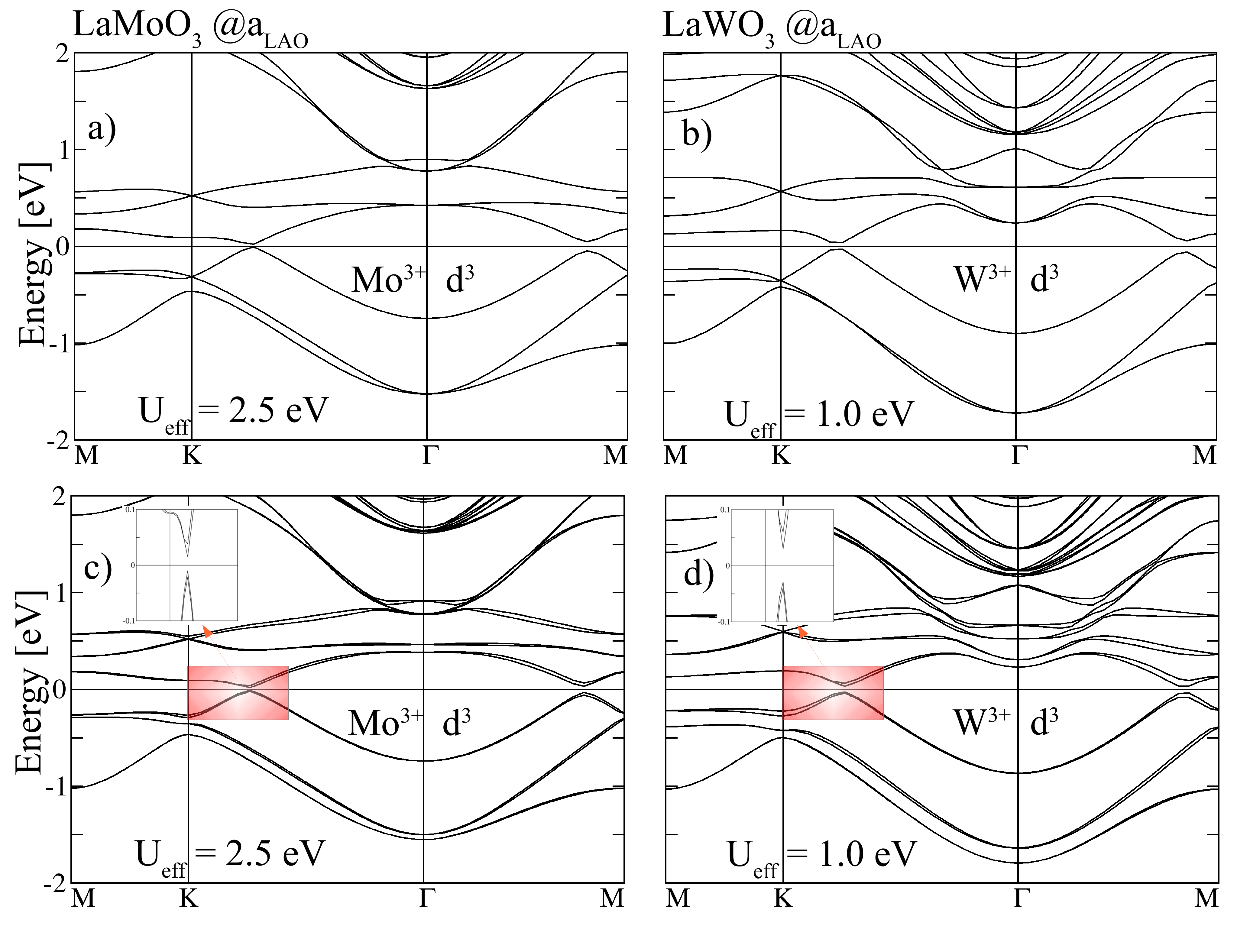}
\caption{Non-magnetic band structures without (a-b) and with SOC (c-d) of the buckled bilayers $X$\,=\,Mo and W for $U_{\rm eff}$\,=\,2.5 eV and 1.0 eV with the in-plane lattice constants $a$ at $a_{\rm LAO}$, respectively. The Fermi level is set to zero.}
\label{fig:Mo_W_LAO}
\end{figure}

\begin{figure*} [t!]
\includegraphics[width=12cm,keepaspectratio]{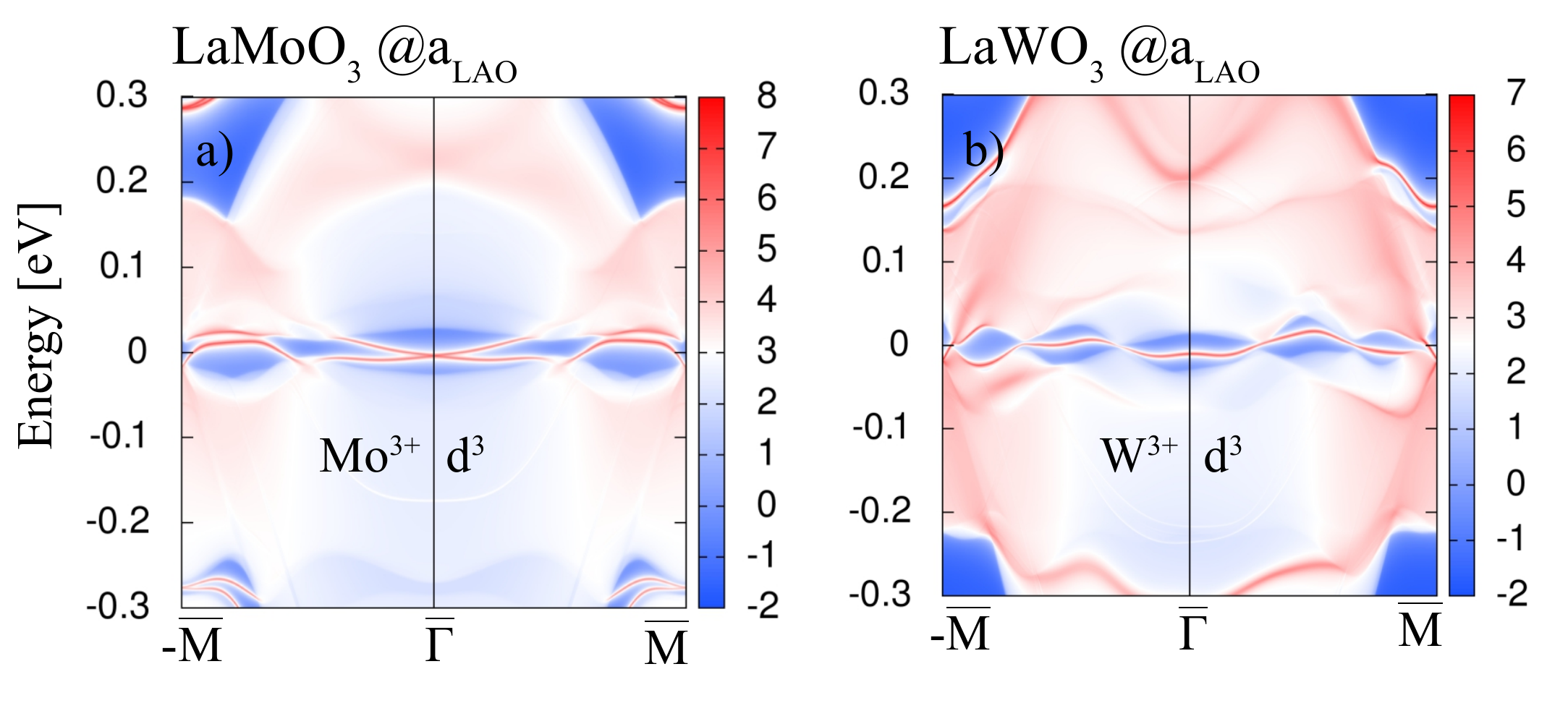}
\caption{The edge states of $X$\,=\,Mo,\,W superlattices shown in (a-b) for (100) surfaces with same color coding as in Fig. \ref{fig:Tc_Pt_BC_edge_states}.}
\label{fig:Edge_states_and_WCCs_Mo_W}
\end{figure*}

\begin{figure} [htbp!]
\includegraphics[width=9.2cm,keepaspectratio]{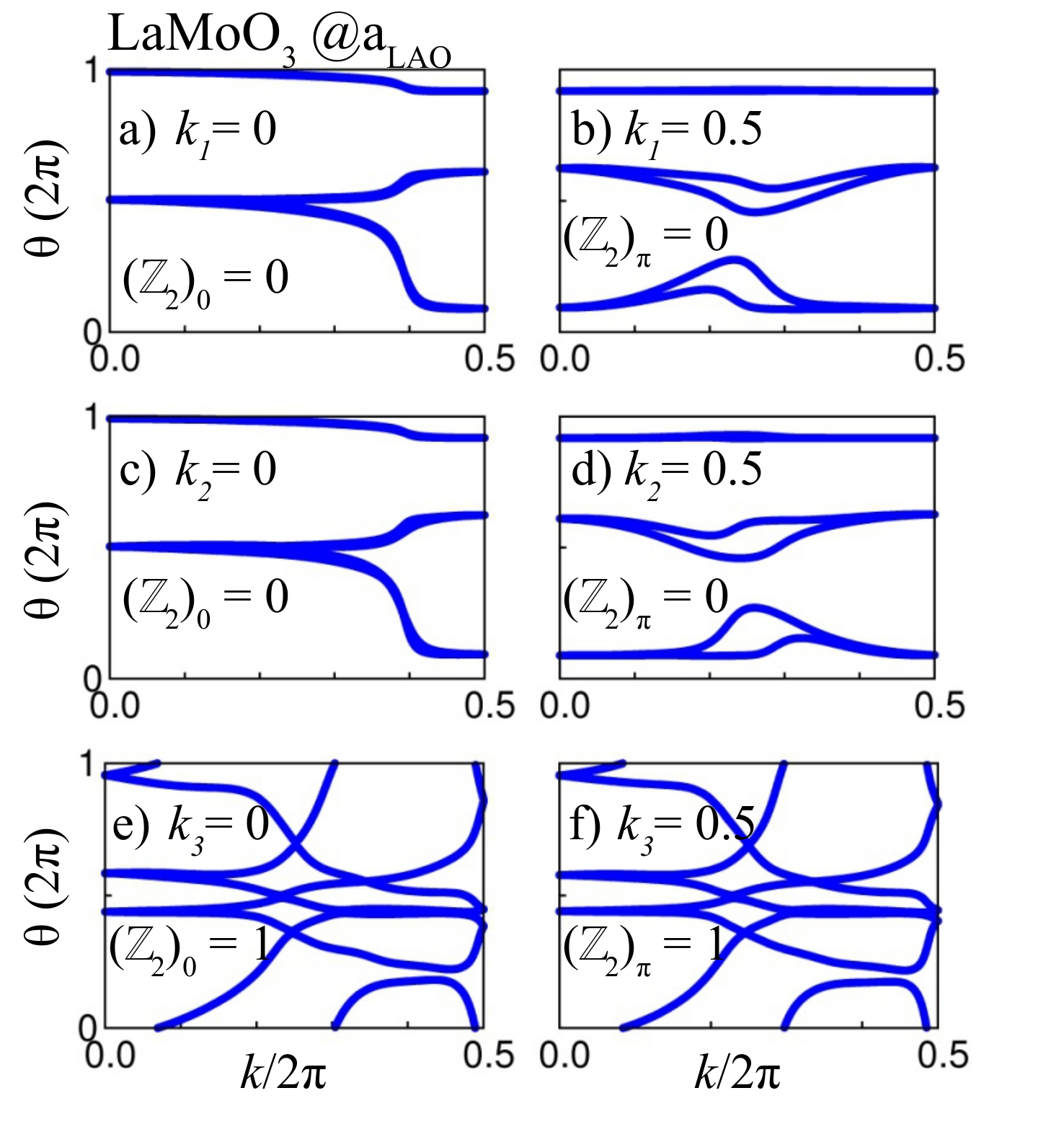}
\caption{The evolution of the Wannier charge centers (WCCs) on six time-reversal invariant momentum planes is shown in (a-f) for $X$\,=\,Mo.}
\label{fig:Mo_Wcc_LAO}
\end{figure}

Besides the potential CI phases in $d^7$ and $d^4$ systems studied above, we investigate the non-magnetic phases of the two $d^3$ systems containing the homologous elements $X$\,=\,Mo and W at $a_{\rm LAO}$. 
Despite the AFM ground state  (cf. Table~\ref{tab:latparam_bilayer_Mo_W}), the potentially interesting systems were identified to be non-magnetic. We note that previous theoretical studies \cite{Michalsky2017} suggest that bulk LaMoO$_3$ and LaWO$_3$ should be non-magnetic. The band structures in Figures \ref{fig:Mo_W_LAO}a and b reveal that the bandwidth of the \tg\ manifold amounts to $\sim1.8$ eV for $X$\,=\,W and $\sim1.5$ eV for $X$\,=\,Mo. The larger bandwidth correlates with the larger extension of the $5d$ orbitals as compared to $4d$. In contrast to bulk LaMoO$_3$ and LaWO$_3$ \cite{Wolfram_Ellialtioglu_2006} which are metallic, the non-magnetic perovskite superlattices exhibit semiconducting behavior with gaps of 28 meV and 62 meV.  When SOC is taken into account both $X$\,=\,Mo and W exhibit topologically non-trivial gaps of 26 meV and 60 meV, respectively. In particular, the degeneracy of bands is lifted along K-$\Gamma$ (cf. Figures \ref{fig:Mo_W_LAO}c and d). 


Since the investigated systems have crystal IS and TRS, $Z_{2}$ can be calculated as a product of parities of all occupied states at the TRIM points by applying the criterion of Fu and Kane \cite{Fu-Kane2007}. In Fig. \ref{fig:Mo_Wcc_LAO}a-f the Wannier function center evolution (WCC) is calculated for $X$\,=\,Mo using the Wilson loop method \cite{Wilson1974,Yu2011} . For the $k_{1}$ and $k_{2}$ planes the $Z_{2}$ indices are 0 whereas for $k_{3}$\,=\,0 and $k_{3}$\,=\,0.5 the $Z_{2}$ indices yield 1.
In order to confirm the topological features of these two systems, we carry out edge state calculations by constructing the MLWFs. The edge Green's function and the local density of states (LDOS) can be simulated  using an iterative method \cite{wanniertools,Sancho1984,Sancho1985}. In the case of $X$\,=\,Mo, one can clearly see a gapless Dirac cone at the $\Gamma$ point (cf. Fig. \ref{fig:Edge_states_and_WCCs_Mo_W}a) whereas one topologically protected chiral edge state is obtained for $X$\,=\,W (cf. Fig. \ref{fig:Edge_states_and_WCCs_Mo_W}b) connecting the valence and conduction bands.

\section{Summary}
\label{sec:fin}
In summary, we investigated the possibility to realise topologically nontrivial states in (111)-oriented perovskite-derived honeycomb La$X$O$_3$ layers with $X=$ $4d$ and $5d$, separated by the band insulator LaAlO$_3$.  The metastable ferromagnetic phases of (LaTcO$_3$)$_2$/(LaAlO$_3$)$_4$(111) and (LaPtO$_3$)$_2$/(LaAlO$_3$)$_4$(111) emerge as CI with $C$\,=\,2 and 1, respectively,  at the lateral lattice constant of LaAlO$_3$ (3.79~\AA). Thereby, the persistence of P321 symmetry, lattice strain and the inclusion of a realistic  Hubbard $U$ term turn out to be crucial. For $X$\,=\,Pd the CI phase appears for $U$ beyond 3.5 eV that likely exceeds the realistic range for a $4d$ compound. For $X$\,=\,Pt the Chern number is reversed beyond $U_{\rm eff}$\,=\,2.0 eV. The CI phase for (LaTcO$_3$)$_2$/(LaAlO$_3$)$_4$(111) is further stabilized under tensile strain (at the lateral lattice constant of LaNiO$_3$), similar to the corundum-based SL (Tc$_{2}$O$_{3}$)$_{1}$/(Al$_{2}$O$_{3}$)$_{5}$(0001). In contrast, for $X$\,=\,Pd and Pt in \lxolao(111) tensile strain lifts the P321 symmetry and induces a site-disproportionation on the two sublattice which opens a trivial band gap and bears analogies to the behavior of the isoelectronic nickelate superlattices\cite{Boris2011,Freeland2011,BlancaRomero2011,Geisler2018,Doennig2014,Doennig2016}.  Further insight into the topological aspects is gained by analyzing the Berry curvatures, edge states and spin textures. A closer inspection of the spin texture for LaPtO$_3$ reveals a spin orientation reversal along the loop of band inversion around K of two bands with opposite spin character.
Moreover, we explored non-magnetic perovskite SLs where TRS is preserved and identified $X$\,=\,Mo and W as potential candidates for $Z_{2}$ TIs. The existence of edge states and non-trivial $Z_{2}$ indices supports this outcome. Recent experimental studies reported the succesfull growth of (111)-oriented nickelates \cite{Middey2012,Middey2016,Keimer2018} as well as nickelate and manganate perovskite heterostructures \cite{Gibert2012}. Thus, we trust that our theoretical predictions will encourage further experimental efforts to synthesize and characterize the proposed systems.

\begin{acknowledgments}
We acknowledge discussions with Warren E. Pickett and D. Khomskii on related systems and funding by the German Science Foundation within CRC/TRR80, project G03 and computational time at the Leibniz Rechenzentrum, project pr87ro.
\end{acknowledgments}

\section*{Contributions}
O.K. performed the calculations under the guidance of R.P., O.K. and R.P. analyzed and interpreted the results and wrote the manuscript. 

\section*{Competing interests}
The authors declare no competing interests.

\end{document}